# Mise en place de scénarios pour la conception d'outils en Chirurgie Minimalement Invasive

**Guillaume Thomann**
*Laboratoire des Sciences pour la Conception, l'Optimisation et la Production de Grenoble (G-SCOP)*
**Jean Caelen**
*Laboratoire d'Informatique de Grenoble (LIG)*
**Morgan Verdier**
*Laboratoire des Sciences pour la Conception, l'Optimisation et la Production de Grenoble (G-SCOP)*
**Brigitte Meillon**
*Laboratoire d'Informatique de Grenoble (LIG)*

## 1 Introduction et contexte de travail

Les progrès scientifiques et techniques des dernières décennies offrent de plus en plus de possibilités pour satisfaire les besoins des chirurgiens en matériel chirurgical et plus précisément en outils chirurgicaux accompagnés d'outils de visualisation ou de suivi. C'est ainsi que la Chirurgie Minimalement Invasive (Minimally Invasive Surgery ou MIS) a fait son apparition en bloc opératoire dans les années 1990. La MIS a pour objectif principal de rendre les effets postopératoires moins douloureux pour le patient essentiellement en termes de récupération des traumatismes subis. Cela est rendu possible principalement en modifiant les procédures opératoires dans le but d'introduire à l'intérieur du corps humain des outils miniaturisés ou modifiés pour être adaptés à de toutes petites incisions et dans le but de réduire la durée d'intervention et donc le temps d'anesthésie et les traumatismes infligés au patient.

L'apparition sur le marché de besoins spécifiques et la demande de la part des chirurgiens, d'outils chirurgicaux dédiés à ce type d'opérations minimalement invasives ne cesse de croître : cela crée un marché ciblé pour des produits innovants de haute technologie.

Les outils chirurgicaux sont aujourd'hui conçus suite à des discussions informelles (ou entretiens dirigés) entre chirurgiens et concepteurs (ou des représentants des concepteurs). Les termes médicaux utilisés couramment par le corps médical et employés également lors de l'expression de leurs besoins ne permettent généralement pas une compréhension immédiate et claire de la part des concepteurs industriels. De plus les chirurgiens ont un savoir-faire qui est difficilement communicable en raison de la longue expérience et des pratiques spécifiques qu'ils ont acquises. Ceci peut engendrer un dysfonctionnement dans le cycle de définition du produit par les concepteurs.

Ce constat impose une réflexion plus approfondie concernant la méthode utilisée pour concevoir des outils chirurgicaux. Nous voulons, en accord avec les chirurgiens, réfléchir et finalement proposer une nouvelle organisation du processus de conception d'outils dédiés aux opérations minimalement invasives de type *arthrodèse lombaire* (fracture de la colonne vertébrale). C'est dans ce cadre de la conception de nouveaux outils chirurgicaux dédiés à ces opérations, que nous réfléchissons à un processus de conception innovant pour la réalisation d'ancillaire adapté à la MIS.

Dans la section suivante, nous décrivons ce contexte médical particulier dans lequel nous travaillons. Pour la conception de nouveaux outils, nous avons l'ambition de placer l'utilisateur – le chirurgien – au centre du processus de conception des outils, non plus pour concevoir *pour* lui, mais pour concevoir *avec* lui. Pour cela, dans la troisième section, nous nous appuyons sur la notion de « conception centrée utilisateur » et sur la notion de *scénario*. Les travaux décrits ci-après consistent donc à appliquer les méthodologies de conception, développées initialement dans le domaine de l'informatique, au monde chirurgical. Dans les deux dernières parties, nous détaillons étape par étape notre démarche expérimentale.

**2 Contexte de travail : les opérations d'arthrodèse lombaire**

Le docteur JT (chef d'unité fonctionnelle de traumatologie au CHU de Grenoble) est spécialiste des opérations de chirurgie traumatique ; dans ce cadre, il effectue régulièrement des opérations d'arthrodèse lombaires. Celles-ci consistent à remettre en état la colonne vertébrale d'un patient ayant subi un grave traumatisme (fracture, choc, etc.) au niveau des vertèbres lombaires (suite à une chute de parapente, de moto, de ski, etc., situations très fréquentes dans les régions de montagne).

Cette intervention est très traumatisante pour le patient et les contraintes postopératoires sont très douloureuses (notamment à cause du décollement des muscles vertébraux) et handicapantes (anesthésie totale de longue durée, long séjour à l'hôpital, nombreuses semaines d'arrêt de travail, etc.). Cette opération nécessite en effet une incision conséquente (25 centimètres de long) dans le bas du dos du patient et la mise en place d'un appareillage lourd de redressement de la colonne vertébrale (figures 1).

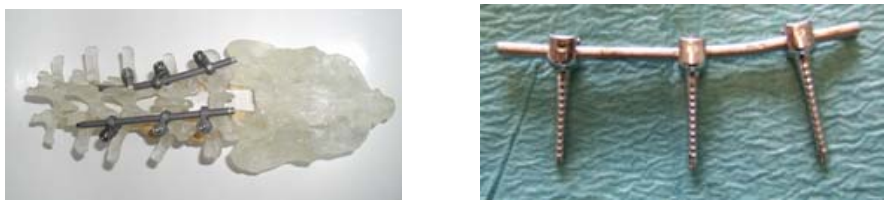

**Figures 1.** *Ancillaire mis en place sur la colonne vertébrale du patient lors de l'opération d'arthrodèse lombaire.*

Pour réduire ces conséquences postopératoires handicapantes, le docteur JT a exprimé le besoin de disposer d'outils chirurgicaux mini-invasifs. Ceux-ci devraient permettre de réaliser l'opération d'arthrodèse lombaire non pas en faisant de grandes incisions, mais en venant placer l'outillage nécessaire à l'intervention par de petits entailles judicieusement positionnées au bas du dos du patient. Il pose comme contrainte d'utiliser une technique nouvelle mais pas trop innovante, c'est-à-dire s'appuyant sur des gestes déjà acquis et des outils dérivés d'autres types d'opérations de type MIS. En effet, pour la sécurité du patient il est préférable de procéder pas à pas dans l'introduction de nouvelles pratiques.

Cette situation constitue donc pour nous une opportunité, un point d'entrée et un cadre concret de recherche pour les analyses du processus de conception de matériel chirurgical innovant. La tâche du chirurgien est relativement « physique », les gestes doivent être précis (notamment pour visser des broches dans les vertèbres sans atteindre la moelle épinière) et en même temps énergiques et énergétiques. Les efforts sont relativement importants pour pénétrer dans l'os, fixer les barres, redresser la colonne (des leviers sont nécessaires). Cela doit se faire sans trop de préparation car certains patients doivent être opérés d'urgence. La procédure doit devenir très routinière et les gestes doivent être faits de manière quasi-automatique pour ne pas dépendre du degré de stress ou de fatigue du chirurgien (qui doit pouvoir opérer à toute heure du jour ou de la nuit).

On le voit donc, ce type d'opération se prête mal à une description verbale car les actes et les gestes passent par des « ressentis », des retours d'effort proprioceptifs et des représentations cognitives fondées sur la mémoire des gestes, des représentations anatomiques imagées et des cas d'expérience accumulés. La question de l'élicitation de ces paramètres est donc au centre de notre problématique. Nous partons du principe que le chirurgien est le mieux placé pour faire émerger ses besoins, non pas nécessairement de manière verbale mais par une mise en situation expérimentale gestuelle.

Nous allons maintenant introduire la notion de conception centrée utilisateur qui peut nous aider à la prise en compte de ces besoins.

**3 Qu'est-ce que la « Conception Centrée Utilisateur » ?**

Dès la naissance de la terminologie de « conception centrée utilisateur » (UCD en anglais) dans le domaine de l'informatique dans les années 1980, l'objectif d'un tel processus était de prendre en compte l'utilisateur dans la conception d'une Interface Homme-Machine (IHM). Cette approche place non seulement l'utilisateur, mais aussi la tâche qu'il doit effectuer (dans la condition où la tâche est correctement définie) au centre du processus de conception [NOR 86].

L'UCD se concentre sur les facteurs cognitifs (tels que la perception, la mémoire, l'apprentissage, la résolution de problèmes, etc.) et sur la manière dont ils interviennent pendant les interactions personne-objet [KAT 98]. Par exemple, lors du développement d'interfaces Web, l'auteur établit que cette méthode aide à répondre aux questions des utilisateurs, à la définition de leurs tâches et de leurs objectifs et permet ainsi d'aboutir à des spécifications respectant les recommandations ergonomiques. Grâce à cette méthode, nous pourrons répondre aux questions suivantes :

– quelles sont les tâches et les objectifs des utilisateurs ? (en termes perceptifs et cognitifs),

– quels sont les niveaux d'expérience (et de performance) des utilisateurs avec ce produit ou des produits annexes ?

– quelles fonctions doit apporter ce produit aux utilisateurs ?

– de quelles informations peuvent avoir besoin les utilisateurs et sous quelle forme ?

– comment les utilisateurs pensent-ils que ce produit devrait fonctionner ?

L'UCD et l'intérêt de sa mise en place ont déjà été discutés dans de nombreux travaux liés au développement d'outils et/ou de logiciels se rapportant aux IHM [GRU 02]. Dans une perspective efficace de conception de produit toujours plus utile (et aussi plus rentable), des méthodologies parallèles à l'UCD sont apparues. Par exemple, dans le domaine de l'Environnement Informatique pour l'Apprentissage Humain (EIAH), les informaticiens sont souvent amenés à travailler avec des partenaires des SHS (Sciences Humaines et Sociales) [JEA 04]. Ces projets, nécessairement pluridisciplinaires, ne peuvent être conduits sans faire appel à des experts ou chercheurs en sciences de l'éducation, à des enseignants et à des apprenants, en sus des informaticiens. Il faut donc que la méthodologie de recherche donne une place centrale à la collaboration entre enseignants, chercheurs et techniciens [COL 96]. Pour le développement de logiciels éducatifs par exemple, il est indispensable d'intégrer des enseignants à la conception des systèmes qu'ils vont utiliser ensuite.

Pour renforcer ce point de vue et l'élargir au-delà de l'informatique, nous pouvons citer une expérience de conception d'équipement dédié à la préparation de nourriture au Bénin [GOD 06]. Ce travail décrit la volonté d'intégrer l'utilisateur et de mettre en avant sa participation lors de la conception de l'outil. L'interaction entre utilisateurs et concepteurs est mise en évidence dans le but d'aboutir à la conception d'un équipement répondant efficacement à leurs exigences réciproques. Lors de cette étude, il s'agit du *plus facile à utiliser* et du *plus facile à manipuler*. Lorsque la tâche de l'utilisateur n'est pas totalement définie, la conception centrée utilisateur ne suffit plus. Pour définir plus précisément cette tâche, il est nécessaire de faire appel aux utilisateurs non plus seulement pour étudier leur comportement face à une situation donnée, mais en tant que concepteur. Il s'agit alors de « conception participative » (Participatory Design ou PD). La démarche PD propose d'associer les utilisateurs au processus de conception dès le début du projet, en partant du principe qu'ils savent (ou qu'ils sont capables de découvrir) ce dont ils ont besoin, et qu'ils peuvent aussi avoir des idées novatrices [GRE 91]. Dans notre étude, la finalité de l'opération est connue, mais le processus opératoire du chirurgien et les outils à employer pour réaliser l'opération chirurgicale sont à proposer.

On distingue la notion de « conception informative » [JEA 04] de celle de PD représentées sur le graphique ci-dessous (figure 2) qui schématise les degrés d'implication des utilisateurs et des chercheurs dans l'équipe de conception du produit.

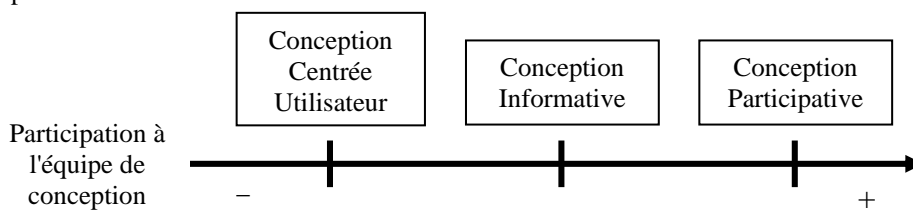

**Figure 2.** *Différentes méthodes de conception suivant le degré de participation des membres de l'équipe de conception.*

Lors de la mise en place de nouvelles méthodes de conception participative, il y a lieu d'expliciter les connaissances et les pratiques des intervenants dans le processus de conception. Ces connaissances sont souvent de l'ordre du savoir-faire chez les ingénieurs ou encore chez les médecins ou les chirurgiens, les pratiques étant issues d'une spécialité sur le terrain. Cette méthodologie oblige donc également à affiner la notion d'utilisateur pour le faire participer à la conception au moment opportun [CAE 05]. Un des grands problèmes que pose la conception participative concerne « la place des utilisateurs et comment les prendre en compte dans le processus de conception ? (À quels moments sont-ils les plus utiles ?) ».

Il nous semble, au vu de ces quelques arguments, que la méthodologie dite PD est un bon point de départ en ce qui concerne la conception de nouveaux outils chirurgicaux pour les opérations d'arthrodèse lombaire. Puisque nous avons comme objectif de concevoir non seulement les outils innovants, mais également le processus opératoire correspondant, il est indispensable que le chirurgien intervienne non pas uniquement en tant qu'utilisateur, mais également comme concepteur, avec les autres experts et les chercheurs. Pour pousser plus loin l'application de cette méthode, nous allons présenter et analyser ci-dessous la notion de « scénario » pour ensuite essayer de définir un espace d'expérimentation en phase avec nos objectifs.

**4 Vers la mise en place de Scénarios**

Un *scénario* est une façon de découvrir des solutions à des problèmes par des mises en situation des usagers eux-mêmes. Il aide les concepteurs à créer des liens avec les activités des usagers, leurs besoins et leurs attentes [LEM 04]. Le terme de scénario est défini comme : *une personne qui réalise un tâche dans un certain contexte,* ce qui est l'essence même de l'approche centrée utilisateur qui vise à observer une personne en situation de travail. Il s'agit d'une des manières parmi les plus efficaces pour cerner la séquence des actions faites par les usagers et établir un modèle d'activité. Les scénarios que nous voulons ainsi mettre en place ont aussi pour objectif de concrétiser de nouveaux concepts ou de nouvelles idées.

Pour proposer et mettre en place des scénarios, des éléments importants sont à définir [CAR 99]. Ils incluent ou présupposent des caractéristiques générales : par exemple, le scénario raconté par un conteur débute explicitement par un état initial par rapport à l'épisode décrit. Le scénario inclut également des agents et des acteurs à identifier. Chaque agent a des objectifs à atteindre. Il n'est pas rare que les objectifs changent durant le scénario en fonction des circonstances. Tous les scénarios font intervenir au moins un agent et au moins un objectif.

Les scénarios évoluent dans le temps et sont constitués de séquences *d'actions* et *d'événements*. C'est à dire des actions que les agents font, des événements qui leur arrivent, des changements dans les circonstances de ces événements. Ces événements et actions particulières peuvent faciliter, compliquer ou n'avoir aucune conséquence par rapport aux objectifs visés. L'analyse du cheminement emprunté pour satisfaire le ou les *objectifs* du scénario permettra de proposer des solutions de conception en prenant en compte les exigences, les savoir-faire, et encore de nombreuses autres particularités liées à l'utilisateur et à son environnement.

Nous avons une certaine pratique de la conception participative pour des IHM : un exemple d'étude centrée utilisateur avec application à un scénario a été effectuée par le laboratoire CLIPS pour estimer l'utilisabilité d'un agent conversationnel animé Angela [GOL 05]. Il s'agissait de recueillir tout au long de l'expérimentation, les remarques et commentaires des sujets concernant leur ressenti par rapport à Angela. Un expérimentateur se trouvait avec l'utilisateur pendant toute la séance afin de lui présenter les tâches à effectuer et d'interagir avec lui au besoin. Un observateur prenait des notes sur l'exécution des tâches et sur les commentaires fournis par les utilisateurs. La séance était enregistrée en temps réel pour de futures analyses. Angela se caractérise essentiellement par sa gestuelle (elle montre des zones de l'écran, prend des attitudes et des postures du corps, etc.), ce qui la rapproche quelque peu du cas présent de chirurgie. Nous avons mis au point une base d'annotation de gestes lors de cette expérience.

Nous allons dans les parties suivantes, détailler les travaux préliminaires qui nous permettent maintenant de nous positionner dans le domaine particulier de la chirurgie traumatologique. Il s'agit de nous approprier le domaine de recherche et de nous familiariser avec les terminologies spécifiques dans un premier temps puis de nous imprégner des conditions opératoires dans un second temps.

**5 Travaux préliminaires à la mise en place de scénarios**

Les recherches sur Angela ont démontré la valeur du travail avec un acteur et la mise en place de scénarios pour permettre aux concepteurs de mieux appréhender les besoins liés à des observations de la gestuelle. Nous avons donc décidé de nous orienter vers une même voie expérimentale pour la conception d'outils chirurgicaux innovants. Cependant, avant de proposer aux utilisateurs des scénarios permettant d'utiliser des prototypes d'outils chirurgicaux associés à de nouvelles procédures opératoires, il a fallu nous imprégner des pratiques et des procédures des chirurgiens.

*5.1 Etape 1 : Observer le travail de l'utilisateur*

La première étape a consisté à observer les faits et gestes des spécialistes lors des opérations chirurgicales actuelles. Il s'agissait ici d'acquérir des informations sur les utilisateurs en recueillant des données sur le terrain. De nombreuses heures ont été passées en salle d'opération pour nous approprier la procédure opératoire, les outils chirurgicaux utilisés et surtout la terminologie employée. Cette analyse a été indispensable pour nous familiariser avec le milieu chirurgical et surtout pour comprendre la séquence des actions réalisées par les chirurgiens et partager leur vocabulaire très spécialisé.

De nombreuses discussions jointes aux observations effectuées sur le terrain ont été essentielles. En particulier nous avons constaté que le contexte implique que les actions de l'utilisateur doivent impérativement être effectuées en salle d'opération. Cette remarque paraissant à première vue anodine impliquera pourtant de nombreux choix de conception de produit : l'équipe en bloc opératoire a des contraintes, une disposition organisationnelle précise et des automatismes de fonctionnement rodés voire imposés. Il faudra donc, dans la mesure du possible, que les outils et la procédure opératoire proposée ne modifient pas les relations entre les chirurgiens et leur équipe assistante.

*5.2 Etape 2 : Organisation des informations recueillies*

La seconde étape concerne l'organisation, le tri, la mise en forme et la communication des informations recueillies. Des heures de vidéo ont été tournées, des photos des différentes actions ont été collectées et organisées pour analyser les actions relatant la procédure opératoire suivie par l'utilisateur [VER 06].

Notre objectif étant de réaliser des prototypes d'outils et de les faire tester aux chirurgiens avec le maximum d'hybridation aux pratiques en cours, nous avons retracé la chronologie de la procédure opératoire principale suivie par un arbre de tâches : chaque étape de cette procédure y est détaillée avec le temps nécessaire alloué. Cet arbre nous permet d'avoir une vue d'ensemble de l'opération chirurgicale et des outils associés à chaque étape, ainsi qu'une nomenclature qui nous indique plus précisément le nom des outils et leur fonction. Nous pouvons citer par exemple les outils numérotés 10 et 11 (figures 3 : distracteur et sa pince associée) qui sont utilisés simultanément pour réaliser la distraction des vertèbres en éloignant les têtes de vis l'une de l'autre [VER 06].

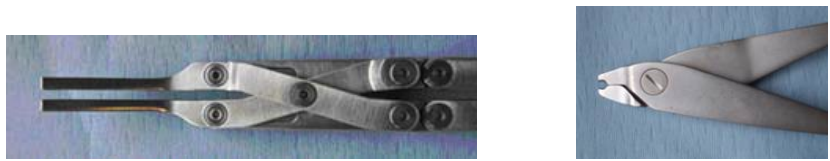

**Figures 3.** *Photographie de deux outils utilisés pour l'étape de distraction lors de l'opération d'arthrodèse lombaire.*

Cette organisation et cette mise en forme des faits relatant l'usage des outils par l'utilisateur nous paraissent indispensables à une bonne compréhension et une bonne représentation de son activité : nous ne pourrons proposer des solutions innovantes qu'en comprenant les besoins du chirurgien et donc en analysant minutieusement son activité – qui rappelons-le est difficilement verbalisable.

### *5.3 Etape 3 : Réflexions vers le nouveau produit*

Le choix de l'organisation de travail et sa mise en place nous ont permis de dégager les étapes opératoires spécifiques liées au caractère traumatisant de l'opération chirurgicale. Cette représentation a ainsi permis une réflexion d'une part sur les modifications à apporter à la procédure opératoire et d'autre part sur la conception de nouveaux outils chirurgicaux. C'est ainsi qu'une nouvelle proposition de modèle de tâche a été faite et discutée avec le chirurgien [VER 06].

Il est ressorti du questionnement proposé par [KAT 98] au paragraphe 3, deux points cruciaux qui doivent apparaître lors de la mise en place des futurs scénarios :

– la prise en compte des fonctionnalités que devront satisfaire les nouveaux outils chirurgicaux,

– la mise en place d'une nouvelle procédure opératoire à suivre par le chirurgien.

### 6 Déroulement des scénarios pour la conception d'outils innovants en MIS

Pendant une opération, des erreurs médicales peuvent se produire et avoir non seulement comme conséquences des problèmes dans le traitement, le suivi ou le diagnostic efficace d'un patient, mais dans certains cas peuvent mener à des traumatismes graves ou définitifs. Il est important que des dispositifs médicaux soient conçus en considérant l'impact de la conception sur une utilisation fiable [SAW 96]. Pour minimiser l'implication des facteurs humains et donc l'utilisation d'outils chirurgicaux par les spécialistes lors des opérations, il est indispensable de tester les prototypes proposés dans l'environnement réel de l'utilisateur. Il faut effectivement que l'utilisateur soit dans les conditions réelles pour qu'un maximum de paramètres soit pris en compte lors des scénarios proposés.

*6.1 Etape 4 : Réalisation du nouveau produit et préparation du scénario*

Pour écrire un scénario, il faut décrire dans un langage simple l'interaction qui a besoin d'être mise en place. Il est important de mettre de coté des références à la technologie, excepté lorsque cette technologie représente une contrainte de conception qui doit être représentée [GAF 00]. Il faut donc toujours faire relire le scénario par un utilisateur pour être sûr qu'il est représentatif du monde dans lequel il évolue. C'est ainsi que suite à de nombreuses discussions avec les chirurgiens, nous avons pu proposer un premier prototype d'outil pour réaliser l'opération en MIS [THO 07]. Cette proposition a été débattue en présence des experts et avec leur accord, il a été décidé de s'inspirer d'un outil utilisé pour des opérations de fracture du fémur, cet outil chirurgical étant déjà utilisé ponctuellement par le chirurgien.

La confection d'un mannequin a été indispensable pour la réalisation d'un premier scénario d'utilisation du prototype d'outil chirurgical. La conception proche de la réalité de ce mannequin a été possible grâce aux indications des spécialistes. Ils nous ont orientés de manière à ce que leurs actions face au mannequin soient le plus proche possible de la réalité : épaisseur de la peau, densité des organes traversés par les outils, fracture de la colonne au niveau de la première vertèbre lombaire, etc.

En ce qui concerne la préparation du scénario, nous avons décidé en accord avec le chirurgien, de nous placer à une certaine étape de la nouvelle procédure opératoire établie. De cette manière, le scénario a permis de tester surtout le prototype d'outil proposé. Cette décision de scénario a impliqué une préparation particulière du mannequin, notamment le pré-positionnement de vis dans les pédicules (figure 4).

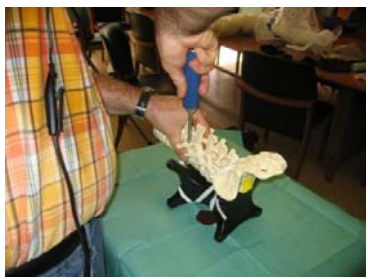

**Figures 4.** *Photographies de la mise en place des vis pédiculaires dans la colonne fracturée*

Les instructions ont été clairement notifiées au chirurgien pour qu'un maximum d'informations soient récupérées au terme de cette expérience. Des caméras frontales et large-champ ont été installées ainsi qu'un micro-cravate pour enregistrer les remarques du chirurgien.

De manière similaire à l'approche nous ayant permis de proposer de nouveaux outils et une procédure opératoire associée, nous nous sommes inspirés de l'étude sur la préparation de scénarios pour organiser notre expérimentation. Nous avons proposé :

– un déroulement contenant un début et une fin (implant en place),

– une « histoire » impliquant le chirurgien, un assistant et les instruments utilisés dans l'environnement habituel de travail.

### *6.2 Etape 5 : Déroulement du scénario*

Le scénario a débuté à l'étape spécifique précisée dans l' « histoire » (vis pédiculaires en place sur la colonne) et s'est terminé une fois l'implant totalement mis en place. Lors de ce scénario, le chirurgien s'est aisément servi de l'appareillage usuel classique mis à sa disposition. Il n'a pas paru gêné par les personnes présentes en bloc opératoire ni par les instruments annexes utilisés pour l'expérimentation (caméras, micro). Sur la figure 5, nous pouvons observer par exemple deux gestes chirurgicaux importants : le repérage de l'épineuse avec le pouce de la main gauche qui permet le geste chirurgical (de la main droite) consistant à inciser la peau du patient à l'endroit de la future insertion de l'outil. Ces gestes chirurgicaux dépendent uniquement de la qualité du mannequin réalisé.

Vers la fin du scénario (figure 6), le prototype d'outil est entièrement en place dans le mannequin, il est tenu par l'assistant du chirurgien. Nous remarquons le chirurgien en train de serrer une des vis sur la tige transversale afin de solidariser cette dernière à la colonne vertébrale. Cette figure nous montre également la présence de la caméra frontale nous renseignant sur l'orientation et la cible du regard de l'utilisateur.

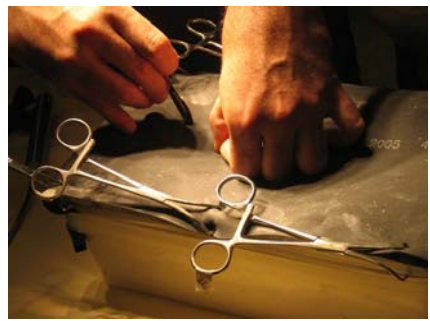 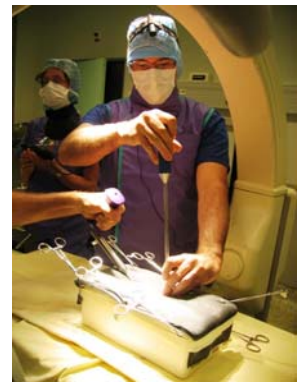

**Figure 5.** *(à gauche) Le repérage de l'épineuse et incision de la peau du patient.*

**Figure 6.** *(à droite) Chirurgien en action vers la fin du scénario.*

### *6.3 Etape 6 : Premiers résultats et améliorations des conditions de déroulement du scénario*

Un dispositif important de collecte d'information ayant été mis en place lors de cette première expérimentation, une grosse partie du travail consiste à analyser et à traiter ces données. Plusieurs améliorations ont été envisagées, principalement à partir du visionnage des vidéos tournées et du discours enregistré pendant le scénario. Deux flux vidéo et un flux audio ont été enregistrés :

– le premier flux vidéo (caméra située sur la tête du chirurgien) nous informe clairement sur l'attention que porte le chirurgien sur l'ensemble de son environnement et sur la succession des événements se produisant dans le bloc opératoire,

– le second (caméra filmant l'ensemble de la scène) nous permet de comprendre l'ensemble de l'organisation dans la salle, autour de l'utilisateur,

– l'enregistrement de la voix du chirurgien (micro cravate) est indispensable pour une analyse a posteriori de ses commentaires.

Il est indispensable de souligner l'importance donnée à la parole du spécialiste lors du déroulement du scénario. Effectivement, le discours de l'utilisateur doit être aisément identifiable par les concepteurs : une remarque critique, une marque de satisfaction ou encore un conseil sur l'utilisation du prototype en cours d'expérimentation lors du scénario doivent absolument être identifiés et mis en avant pour l'évolution future du produit. Ce tri (ou filtrage) des événements ayant eu lieu lors du scénario est actuellement en cours de traitement grâce au logiciel *The Observer XT* disponible au laboratoire.

Suite à ce premier scénario, nous avons pu mettre en avant des marques de satisfaction et des paramètres à modifier pour le rendre plus efficace :

– la position des caméras et du micro cravate sont convenables et seront certainement à conserver pour de prochaines expérimentations,

– nous avons détecté (d'après les sensations exprimées oralement par l'utilisateur) des mauvais choix de matériaux pour la conception du mannequin, notamment en ce qui concerne la matière représentant les muscles dans le dos du patient,

– les outils conçus sont compatibles avec le matériel de guidage utilisé lors des opérations chirurgicales (radiographies pour du repérage spatial),

– le premier prototype d'outil conçu répond aux principaux besoins exprimés par l'utilisateur, mais de nouvelles fonctionnalités doivent y être ajoutées et il devra être utilisé en complément d'autres outils lors des futurs scénarios qui seront proposés.

**7 Conclusion**

Lors du présent travail de recherche, nous nous sommes focalisés sur l'aide à la conception d'un ancillaire destiné à une opération en Chirurgie Minimalement Invasive. Les difficultés qu'ont les concepteurs pour répondre le plus exactement possible aux besoins des utilisateurs nous ont poussé à réfléchir à une méthodologie de conception qui permette de contrecarrer ce handicap.

C'est pour cela que nous nous sommes tout naturellement dirigés vers l'UCD pour pouvoir, à terme, proposer aux concepteurs, une méthode moins gourmande en temps et plus efficace. Une étude bibliographique de l'UCD nous a informés que cette méthode était déjà très utilisée dans le domaine informatique, mais aussi qu'elle nous menait tout naturellement vers les notions de Pd et de SBD.

C'est donc en nous basant sur des exemples que nous avons, dans cette étude, proposé une réflexion centrée sur le chirurgien qui nous a ensuite permis d'expérimenter un scénario autour de ce dernier. Grâce à ce scénario, nous avons testé une partie de la nouvelle procédure opératoire et évalué un des prototypes d'outil chirurgical adapté aux nouvelles exigences de la MIS. Les moyens mis en œuvre (notamment audio et vidéo) nous permettent actuellement une réflexion et des prises de décision *a posteriori* pour apporter des améliorations lors des prochaines expérimentations sur :

– le mannequin confectionné,

– la procédure opératoire proposée,

– les prototypes d'outils chirurgicaux mis à disposition de l'utilisateur.

Le but du travail actuel et futur est donc clairement d'intégrer plus efficacement les besoins et exigences des utilisateurs dans le processus de conception d'outils chirurgicaux, notamment par la création de conditions expérimentales adaptées. Il s'agit également maintenant de réaliser à nouveau plusieurs scénarios successifs dans le but de proposer un modèle co-évolutif du processus de conception intégrant les prototypes proposés et les utilisations.

# Bibliographie

Recherche MEI (Mécanique, Energétique, Ingénierie), option MCGM (Mécanique : Conception-Géomécanique-Matériaux), Grenoble, 2006.